\documentclass[
9pt,
twocolumn,
twoside,
]{opticajnl}
\journal{ol} 
\setboolean{shortarticle}{true}

\usepackage{lineno}

\usepackage{graphicx} 
\usepackage{dcolumn} 
\usepackage{bm} 
\usepackage{braket}
\usepackage{mathptmx}
\usepackage{siunitx}
\usepackage[utf8]{inputenc}
\usepackage[T1]{fontenc}

\newcommand{\qty}[2]{\SI{#1}{#2}}

\title{Two-dimensional electronic spectroscopy of an ultracold gas}

\author[1]{Friedemann Landmesser}
\author[1]{Tobias Sixt}
\author[1,2,*]{Katrin Dulitz}
\author[1,*]{Lukas Bruder}
\author[1]{Frank Stienkemeier}

\affil[1]{Institute of Physics, University of Freiburg, Hermann-Herder-Stra{\ss}e 3, 79104 Freiburg, Germany}
\affil[2]{Institut für Ionenphysik und Angewandte Physik, Universität Innsbruck, 6020 Innsbruck, Austria}
\affil[*]{Corresponding authors: 
katrin.erath-dulitz@uibk.ac.at, lukas.bruder@physik.uni-freiburg.de}

\begin{abstract}
Femtosecond coherent multidimensional spectroscopy is demonstrated for an ultracold gas.
For this, a setup for phase modulation spectroscopy is used to probe the $3^2\mathrm{S}_{1/2} - 2^2\mathrm{P}_{1/2, 3/2}$ transition in an 800\,µK-cold sample of $^7$Li atoms confined in a magneto-optical trap.
The observation of a double quantum coherence response, a signature of interparticle interactions, paves the way for detailed investigations of few- and many-body effects in ultracold atomic and molecular gases using this technique.
The experiment combines a frequency resolution of 3\,GHz with a potential time resolution of 200\,fs, which allows for high-resolution studies of ultracold atoms and molecules both in the frequency and in the time domain.
\end{abstract}

\setboolean{displaycopyright}{true}

\begin{document}

\maketitle

Ultracold atoms and molecules offer numerous exciting research prospects in physics and chemistry, including the modelling of condensed matter systems~\cite{Baranov2012}, precision measurements of fundamental physics~\cite{Safronova2018}, as well as the study of atomic and molecular interactions and dynamics in a regime dominated by quantum effects~\cite{Quemener2012, Heazlewood2021}.
The presence of external fields may even be used to tune the interparticle interaction strength~\cite{Krems2008}.

The photoexcitation and photoionization of ultracold atoms provides access to research fields including
ultracold Rydberg gases~\cite{Browaeys2020} 
and ultracold plasmas~\cite{Killian2007}.
Such experiments are rarely done using femtosecond (fs) lasers, although they deliver coherent, intense laser fields and provide access to the real-time dynamics of the excited system.
The direct observation of electronic dephasing in an ultracold Rydberg gas, induced by an ultrashort laser pulse, provides a prominent recent example for the capabilities of fs lasers in the study of many-body effects~\cite{Takei2016}.
Other recent work on the fs laser excitation of a Bose-Einstein condensate to an ultracold microplasma has also allowed for the monitoring of ultrafast electron cooling~\cite{Kroker2021}.

In general, these studies can strongly benefit from spectroscopic techniques with both high frequency and time resolution as provided by two-dimensional electronic spectroscopy (2DES) in the fs regime~\cite{jonas2003}.
Here, the time evolution of electronic coherences is tracked interferometrically with fs pulses. 
The Fourier spectrum of this signal directly relates to the absorption/emission spectrum of the sample.
Further, 2DES can be combined with the phase modulation (PM) technique~\cite{Tekavec2007} enabling coherent pathway selectivity of incoherent observables.
The high sensitivity of this approach is beneficial for systems with low target densities which 
has recently enabled 2DES studies of cold molecular targets~\cite{Bruder2018a, bangert2022}. 
Moreover, most 2DES methods track the evolution of single-quantum coherences (1QCs), that are electronic coherences induced by one-photon transitions. 
The PM technique enables in addition the tracking of higher-order quantum coherences. 
Such $n$-quantum coherence (nQC) signals provide information about multiphoton transitions and higher lying states, e.g. multi-excitonic states~\cite{turner2010}. 
This property has been used to gain access to
interparticle interactions in thermal atomic vapors, where cooperative effects are otherwise covered by inhomogeneous broadenings~\cite{Bruder2019b, Yu.2019, Yu.2019b, Ames2022, Yu.2022}.

Since ultracold atomic and molecular ensembles with their well-defined structural properties in the quantum regime are ideal testbeds for few- and many-body effects, the combination of ultracold quantum gases with coherent spectroscopy techniques, and their inherent selectivity to weak interparticle couplings, opens up new prospects for detailed studies of such phenomena. 
\begin{figure*} [t]
\centering
\includegraphics[width=17cm]{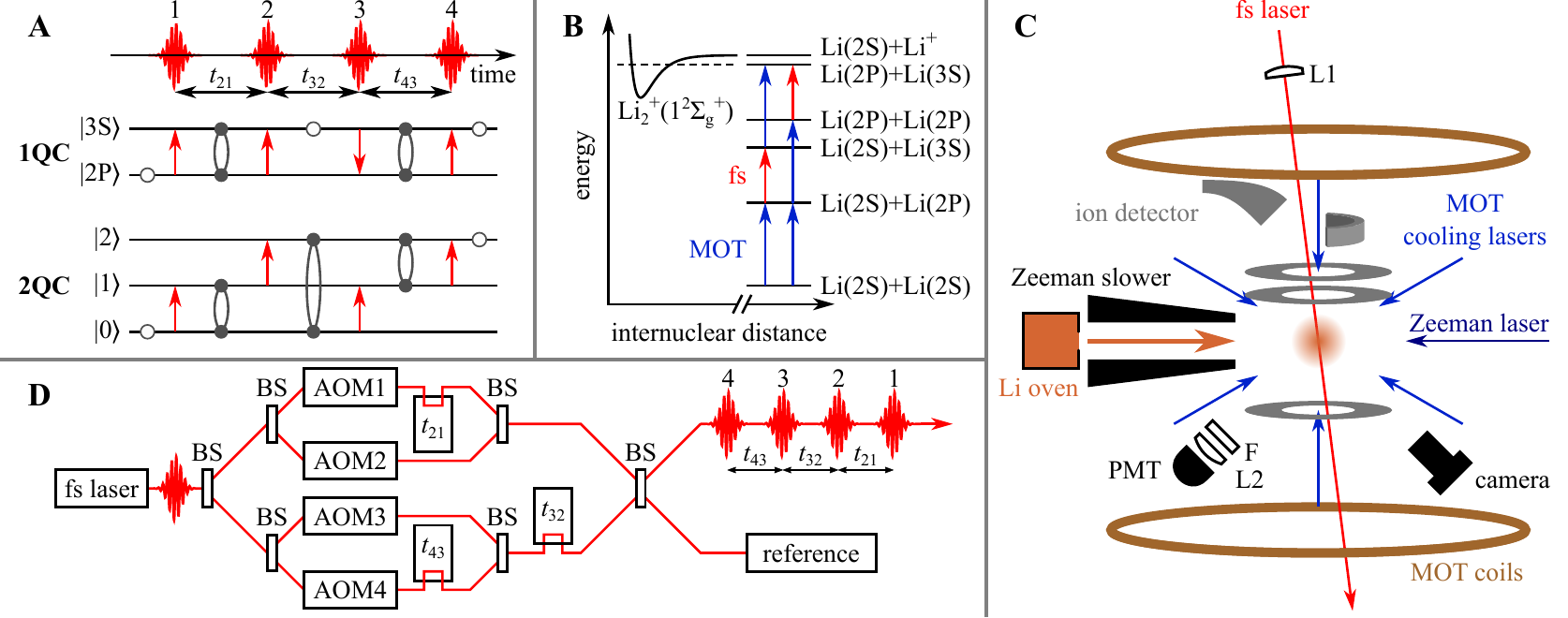}
\caption{
(A) Exemplary coherent fs excitation pathways for 1QC 2DES in a single Li atom and 2QC 2DES for collective states of two Li atoms.
(B) Sketch of ionization scheme following Ref. \cite{Kurz2021a} with asymptotic energy levels of Li$_2$ (not to scale), where a collision of two atoms excited to the $\ket{2\mathrm{P}}$ and $\ket{3\mathrm{S}}$ states can yield a Li$_2^+$ ion.
Transitions excited by the cooling lasers and by the fs laser are indicated by blue and red arrows, respectively.
Experimental setups for the MOT in (C) and for 2DES in (D). 
See main text for details.
Abbreviations: BS: beamsplitter, AOM: acousto-optic modulator, PMT: photomultiplier tube, L1: lens ($f=\qty{400}{\mm}$), F: long-pass filter, L2: lens ($f=\qty{50}{\mm}$). 
}
\label{fig:expsetup}
\end{figure*}
Here, we demonstrate the application of 1QC and 2QC 2DES to an ensemble of ultracold atoms in a magneto-optical trap (MOT).
We choose $^7$Li as a test system, since the interaction of fs radiation with ultracold colliding pairs of Li atoms in a MOT has already been studied in detail~\cite{Kurz2021a}, so that the underlying mechanisms are tangible. 

In the experiment, the $^7$Li atoms are cooled and trapped via the $2^2\mathrm{P}_{3/2} - 2^2\mathrm{S}_{1/2}$ transition (details below). 
The trapped atoms interact with a sequence of four fs laser pulses resonantly driving the $3^2\mathrm{S}_{1/2} \leftarrow 2^2\mathrm{P}_{3/2}$ transition. 
The interaction with the intense fs laser pulses leads to a multitude of linear and nonlinear signals, of which a specific subset (described in Fig.~\ref{fig:expsetup}(A)) is selected and detected in the experiment. 
Light-matter interactions with single-atom states and collective two-atom states (excitonic states) are possible. 
For a more convenient description, we assign the states as follows: $2^2\mathrm{S}_{1/2}$ state as $\ket{2\mathrm{S}}$, $2^2\mathrm{P}_{1/2, 3/2}$ state as $\ket{2\mathrm{P}}$ and $3^2\mathrm{S}_{1/2}$ state as $\ket{3\mathrm{S}}$. 
The two-atom states denoted as $\ket{0} - \ket{2}$ correspond to both atoms in the $\ket{2\mathrm{P}}$ state, one atom promoted to the $\ket{3\mathrm{S}}$ and both atoms promoted to the $\ket{3\mathrm{S}}$ state, respectively. 

In the 1QC 2DES measurement (Fig.~\ref{fig:expsetup}(A)), pulses 1 and 3 induce a coherence between the $\ket{2\mathrm{P}}$ and $\ket{3\mathrm{S}}$ state, while pulses 2 and 4 project the coherence onto the excited/ground state manifold. 
Accordingly, the time evolution of the electronic coherences is probed during the time intervals $t_{21}$ and $t_{43}$, while the system evolves on the excited/ground state manifold during the interval $t_{32}$. 
The probability to end in an eigenstate after the interaction with pulse 4 depends on the coherent dynamics induced by pulses 1-4. 
By detecting the fluorescence or photoionization yield, we map this information onto the respective count rate. 
In the experiment, $t_{21}$ and $t_{43}$ are interferometrically scanned and a Fourier transform of the data w.r.t. these time intervals yields 2D frequency-frequency ($\nu_{21}$-$\nu_{43}$) correlation maps as a parametric function of the time interval $t_{32}$ (see results). 
Conversely in the 2QC 2DES case, $t_{21}$ is constant and the combination of pulses 1 and 2 induces a coherent superposition of states $\ket{0}$ and $\ket{2}$. 
In this case, a Fourier transform w.r.t. $t_{32}$ and $t_{43}$ yields 2D spectra correlating the 2QC frequencies ($\nu_{32}$) to the 1QC frequencies ($\nu_{43}$) in the system.

Fig.~\ref{fig:expsetup}(B) shows the interplay between the cooling laser (MOT laser) and the fs laser. 
The results of previous investigations on $^6$Li suggest that the simultaneous resonant excitation of the $2^2\mathrm{P}_{3/2} \leftarrow 2^2\mathrm{S}_{1/2}$ transition during laser cooling and the near-resonant excitation of the $3^2\mathrm{S}_{1/2} \leftarrow 2^2\mathrm{P}_{3/2}$ transition using fs radiation proceeds via an ionization mechanism in which Li atom pairs are excited into an autoionizing molecular state via a ladder mechanism~\cite{Kurz2021a}.
Two pathways were proposed in which two cooling laser photons and one fs laser photon are involved (Fig.~\ref{fig:expsetup}(B)). 
Both are expected to contribute to the ion yield in the experiment. 

Fig.~\ref{fig:expsetup}(C) and (D) show the MOT setup and the fs laser setup, representing the two main parts of the experiment. 
Our MOT setup was described previously in Refs.~\cite{Strebel2012, Grzesiak2019}. 
Therefore, only a brief summary is given here. 
An ultracold trapped sample of $^7$Li is produced by continuously laser cooling the atoms via the $2^2\mathrm{P}_{3/2} - 2^2\mathrm{S}_{1/2}$ transition at a wavelength of \qty{671}{\nano\meter} in a standard three-dimensional (3D)-MOT.
The cooling lasers are detuned by \qty{-12}{\mega\hertz} from the $2^2\mathrm{P}_{3/2} - 2^2\mathrm{S}_{1/2}$ resonance.
Using a rate equation model of the optical pumping process by the MOT lasers and neglecting the influence of the fs laser, we estimate that the steady-state fraction of the atoms in the $2^2\mathrm{P}_{3/2}$ level is $\approx \qty{29}{\percent}$.
To load the MOT, an effusive Li beam is produced by resistive heating of a natural isotope mixture of Li inside an oven and the $^7$Li atoms contained in the atomic beam are subsequently decelerated inside a Zeeman slower. 
The frequency detuning of the laser for Zeeman slowing is further detuned from resonance, so that this laser does not significantly populate the $2^2\mathrm{P}_{3/2}$ level. 
Using fluorescence imaging of the trapped Li atoms with a charge-coupled device camera, the 3D spatial distribution of the atomic cloud is found to be nearly Gaussian with a prolate spheroidal shape (semiaxes of $\approx \qty{1}{\milli\meter}$ and $\qty{2}{\milli\meter}$, respectively). 
The average number of trapped Li atoms is determined as $\num{7e7}$ resulting in a density of $\approx \qty{6e10}{\per\cm\cubed}$. 
Using thermometry measurements~\cite{Grzesiak2019} under similar conditions, the temperature of the atomic cloud is found to be $\approx \qty{800}{\micro\kelvin}$.

The Li MOT is continuously exposed to fs laser radiation (\qty{80}{\MHz} repetition rate, \qty{0.1}{\nano\joule} pulse energy, central wavelength of \qty{812.3}{\nano\meter}, \qty{6.2}{\nano\meter} FWHM bandwidth) with a time resolution limiting pulse duration of \qty{200}{\fs} (intensity FWHM), that is resonant with the $3^2\mathrm{S}_{1/2} - 2^2\mathrm{P}_{1/2, 3/2}$ transition using the optical setup described below.
The fs pulse train is focused into the Li cloud (lens L1 in Fig.~\ref{fig:expsetup}(C)) resulting in a laser beam diameter of about \qty{100}{\micro\meter} in the interaction volume. 
Incidentally, no significant effect of the fs laser radiation on the total, steady-state number of cold atoms in the trap is observed since its radiative force is much weaker than the trapping force of the MOT.

The principle of the 2DES setup in our laboratory (cf. Fig.~\ref{fig:expsetup}(D)) is described elsewhere~\cite{Bruder2019, Yu.2019} and thus only briefly summarized here. 
Four time-ordered laser pulses ($i = 1, 2, 3, 4$) are created in three nested Mach-Zehnder interferometers and are individually phase-modulated at frequencies $\Omega_i$ using acousto-optic modulators (AOMs).
The interpulse delays $t_{21}$, $t_{32}$, and $t_{43}$ are controlled using mechanical delay stages.
To select only specific sets of nonlinear quantum pathways and effectively suppressing background signals, the detected quasi-continuously modulated signals are demodulated by lock-in detection at specific frequencies \cite{Tekavec2007}.
The AOMs are driven at \qty{\approx 155}{\MHz} with relative detunings such that the demodulation frequencies are $\Omega_\mathrm{RP/NRP} = (\Omega_{4} - \Omega_{3}) \mp (\Omega_{2} - \Omega_{1}) = \qty{8}{\kilo\hertz} \mp \qty{5}{\kilo\hertz}$ for rephasing and non-rephasing 1QC 2DES signals, respectively, and $\Omega_\mathrm{2QC} = (\Omega_{4} - \Omega_{1}) + (\Omega_{3} - \Omega_{2}) = \qty{5}{\kilo\hertz} + \qty{1.35}{\kilo\hertz}$ for the 2QC 2DES signals.
In addition, we track the phase fluctuations in the optical interferometers (denoted as ``reference'' in Fig.~\ref{fig:expsetup}(D)). 
To this end, a portion of the interferometer output is spectrally filtered with a monochromator and fed into the lock-in amplifier for heterodyne detection. 
The phase-synchronous demodulation of the fluorescence/ion signal with the optical reference removes the correlated phase noise from the measurement, providing an efficient passive stabilization of the optical interferometers~\cite{Bruder2015a, Tekavec2007}. 
It also enables the detection of the signal in the rotating frame of the spectrally filtered reference which leads to a down shift of the signal frequencies recorded w.r.t. the coherence time delays. 

The fluorescence and ion yields are detected in parallel and are simultaneously processed with two digital lock-in amplifiers. 
The fluorescence emitted from the $3^2\mathrm{S}_{1/2}$ state is collected using a photomultiplier tube.
An optical longpass filter with a cut-on wavelength of \qty{750}{\nano\metre} is used to avoid the detection of fluorescence and stray light from Li laser cooling.
The ions are continuously accelerated onto a channel electron multiplier (CEM)~\cite{Grzesiak2019}. 
Each signal is amplified with a trans-impedance amplifier, where the gain bandwidth for the CEM signal is limited to \qty{50}{\kilo\hertz} to smooth the electronic pulses for the lock-in detection. 

To calculate the complex-valued Fourier spectra, the measured time-domain data are multiplied with a Gaussian apodization function, zero-padded and 2D Fourier-transformed w.r.t. the scanned time delays $t_{21}$, $t_{43}$ (\qty{0}{\ps} to \qty{30}{\ps}) for the 1QC 2DES measurements and the delays $t_{32}$, $t_{43}$ (\qty{0}{\ps} to \qty{8}{\ps}) for the 2QC 2DES measurements. 
In the 1QC 2DES case, the rephasing and non-rephasing spectra are summed up and the real part is shown which yields information about the absorptive part of the third-order susceptibility \cite{Tekavec2007, Bruder2018a, bangert2022}.
For the 2QC 2DES signals, absolute values of the Fourier transforms are plotted, as is common practice \cite{turner2010, Bruder2019b, Yu.2019, Yu.2019b}.
Signal-to-noise ratios (SNRs) are given as peak maximum divided by the root-mean-square of the spectral noise.
To account for the rotating frame detection, the Fourier spectra are up-shifted by the frequency of the reference signal (hence the monochromator frequency)  $\nu_\mathrm{ref} = \qty{369.0187(69)}{\THz}$ for the 1QC frequencies and $2 \nu_\mathrm{ref}$ for the 2QC frequencies, respectively.

\begin{figure} [ht]
\centering
\includegraphics[width=8.5cm]{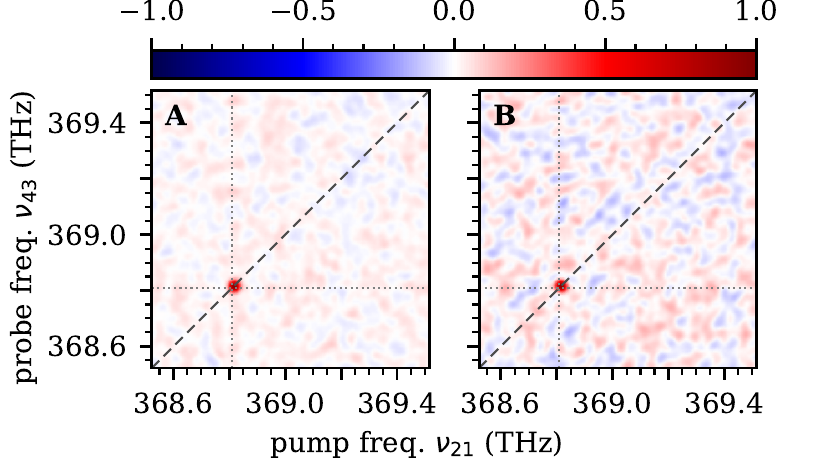}
\caption{
Normalized 1QC 2D correlation spectra (real part) obtained using simultaneous (A) photoion (SNR: 37) and (B) fluorescence (SNR: 20) detection.
A dashed line is used to mark $\nu_{21} = \nu_{43}$. Dotted lines indicate the expected transition frequency $\nu_{3\mathrm{S}_{1/2} - 2\mathrm{P}_{3/2}}$.
}
\label{fig:2des}
\end{figure}

Fig.~\ref{fig:2des} shows 1QC 2D correlation spectra at the evolution time $t_{32} = \qty{0}{\femto \second}$.
A clear peak at the $3^2\text{S}_{1/2} - 2^2\text{P}_{3/2}$ frequency $\nu_{3\mathrm{S}_{1/2} - 2\mathrm{P}_{3/2}} = \qty{368.8083(42)}{\THz}$~\cite{NIST_ASD} is visible in both ion-detected (Fig.~\ref{fig:2des}(A)) and fluorescence-detected (Fig.~\ref{fig:2des}(B)) spectra.
No additional features are observed, as expected. 
A lower SNR is observed for fluorescence detection compared to ion detection. This is due to the significant amount of stray light from the fs laser that reaches the fluorescence detector and due to the limited dynamic range of detection. 

\begin{figure} [ht]
\centering
\includegraphics[width=7.5cm]{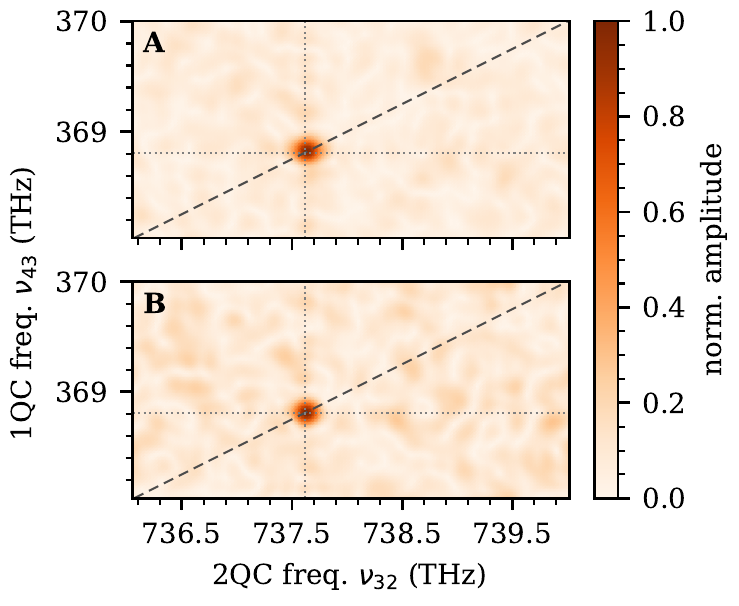}
\caption{
Normalized 2QC 2D correlation spectra obtained using simultaneous (A) photoion (SNR: 13) and (B) fluorescence detection (SNR: 9).
A dashed line is used to mark $\nu_{32} = 2 \nu_{43}$.
Dotted lines indicate the frequencies $\nu_{3\mathrm{S}_{1/2} - 2\mathrm{P}_{3/2}}$ (1QC) and $2 \nu_{3\mathrm{S}_{1/2} - 2\mathrm{P}_{3/2}}$ (2QC).
}
\label{fig:mqc}
\end{figure}

The results of the 2QC 2DES measurements are shown in Fig.~\ref{fig:mqc}. 
Again, a single peak is visible in both the ion and fluorescence channels (Fig.~\ref{fig:mqc}(A) and (B), respectively). As expected, for $\nu_{32}$, the peak appears at the frequency $2(\nu_{3\mathrm{S}_{1/2} - 2\mathrm{P}_{3/2}})$. For $\nu_{43}$, the peak is at $\nu_{3\mathrm{S}_{1/2} - 2\mathrm{P}_{3/2}}$.
This can be understood as a 2QC during $t_{32}$ and a 1QC during $t_{43}$, respectively, as explained above and illustrated in Fig.~\ref{fig:expsetup}(A).
2QC signatures have been previously observed in alkali vapors at similar densities, but at higher temperatures, and they have been attributed to interparticle interactions~\cite{Bruder2019b, Yu.2019, Yu.2019b, Ames2022, Yu.2022}.
We thus interpret the mere existence of the observed 2QC peak as a signature of interacting $^7$Li atoms or atom pairs. 
The high quality of the 1QC and 2QC 2D spectra confirms the feasibility of the presented experimental method. 

\begin{figure} [ht]
\centering
\includegraphics[width=8.cm]{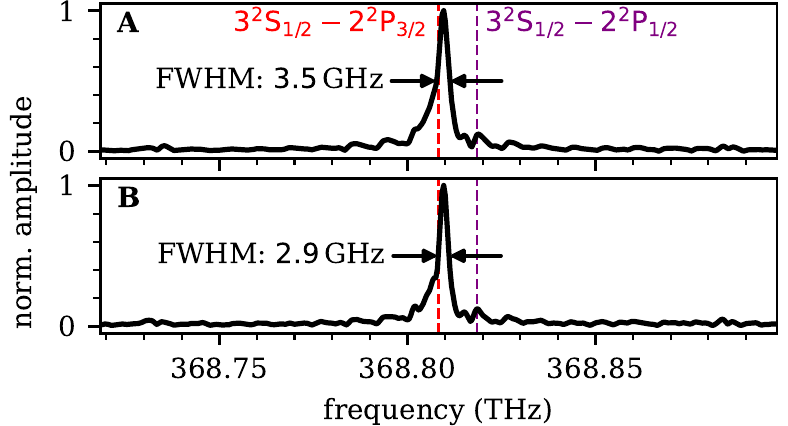}
\caption{
Normalized high-resolution 1D 1QC spectra obtained using (A) photoion (SNR: 200) and (B) fluorescence (SNR: 106) detection.
Transition frequencies~\cite{NIST_ASD} are marked by dashed lines.
}
\label{fig:high_resolution}
\end{figure}

To demonstrate the high-resolution capabilities of this technique, we have also performed one-dimensional (1D) 1QC scans covering also large interpulse delays up to \qty{0.6}{\ns}.
Instead of creating the reference signal from the fs laser using a monochromator, we coupled a continuous-wave laser (\qty{384.0006(1)}{\THz}) into the interferometer to track the interferometer beating and phase fluctuations.
This allows for longer time delays and thus a higher frequency resolution. 
The 1D 1QC spectra for photoion and fluorescence detection are shown in Fig.~\ref{fig:high_resolution}(A) and Fig.~\ref{fig:high_resolution}(B), respectively.
A main peak with a resolution-limited width of $\approx \qty{3}{\giga\hertz}$ full width at half maximum is observed, which we assign to the $3^2\mathrm{S}_{1/2} - 2^2\mathrm{P}_{3/2}$ transition. 
The slightly different linewidths between (A) and (B) and the asymmetric line shapes are due to experimental instabilities occurring during the long acquisition time. 
Taking the refractive index of air \cite{Peck.1972} into account, the peak maximum is blue-shifted by about \qty{1.4}{\giga\hertz} from the literature value, which also has an absolute uncertainty of \qty{4.2}{\giga\hertz}~\cite{NIST_ASD}. 
We also expect a small contribution from the $3^2\mathrm{S}_{1/2} - 2^2\mathrm{P}_{1/2}$ resonance. 
A corresponding peak seems observable although the peak is hardly discernible from the noise. 
Observing this resonance is counterintuitive to the fact, that the cooling laser initially prepares the trapped atoms in the $2^2\mathrm{P}_{3/2}$ state. 
However, spontaneous relaxation or stimulated emission from the $3^2\mathrm{S}_{1/2}$ state into the $2^2\mathrm{P}_{1/2}$ state make this resonance in principle accessible in our experiment.

We have also obtained steady-state ion-time-of-flight spectra to investigate the origin of the detected ions.
To this end, the CEM was operated in an ion counting mode. The extraction electrodes were switched from high voltage to zero for a period of \qty{2}{\micro\second} to allow for an accumulation of ions in the laser interaction volume.
We find that the only contribution to the ion signal intensity stems from Li$_2^+$.
Nevertheless, no spectral evidence for the presence of Li$_2$ molecules in the ultracold sample are observable here.
This suggests that the proposed collisional ionization mechanism~\cite{Kurz2021a} is dominant and that the direct photoionization of ultracold Li atoms is negligible. 
Accordingly, the linewidth of 1QC signals as well as the 2QC signals should contain the time constant of the autoionization process. 
From the observed linewidth in Fig.~\ref{fig:high_resolution}, we deduce that the time scale of the collisional ionization process is $> \qty{100}{ps}$,
which is in accordance with previous reports of slow, ultracold autoionizing collisions, indicating that they are mostly limited by the radiative lifetimes of the involved excited states~\cite{Kurz2021a, Trachy2007}.

In conclusion, we have presented the application of coherent multidimensional electronic spectroscopy to ultracold samples in experiments for the first time.
The feasibility of taking conventional 1QC 2DES spectra was demonstrated by detecting both photoionization and fluorescence observables in parallel.
The simultaneous measurement of different observables may facilitate the extraction of complementary information from 2DES data in the future. 
Furthermore, signatures of two-quantum coherences were observed which highlights the high sensitivity of this technique and provides a new route for the study of interparticle interactions in ultracold ensembles.

This work paves the way for detailed investigations of few- and many-body effects in ultracold atomic and molecular gases using coherent multidimensional electronic spectroscopy.
In recent theory work, Rydberg atoms have been identified as suitable systems for coherent multidimensional spectroscopy ~\cite{Mukherjee2020, Wang2022} owing to the long-range character of their interactions.
Broadband picosecond and fs laser radiation prevents Rydberg blockade effects and thus allows for the study of many-body physics in cold and ultracold atomic and molecular plasmas~\cite{Killian1999, Robinson2000, Morrison2008}, as demonstrated already~\cite{Zhou2014, Takei2016, Sous2019}. 
Yet, broadband excitation of Rydberg gases inevitably leads to a complex response with a large number of contributing Rydberg states. 
Here, 2DES would be of great benefit due to the high frequency resolution despite using broad bandwidth laser pulses. 
In addition, cross-peaks and nQC signals would directly disclose many-body correlations even in highly congested Rydberg spectra. 
Other applications may involve the simultaneous time- and frequency-resolved study of ultrafast dynamics following the photoexcitation of weakly bound ultracold molecules formed by photoassociation, Feshbach resonances or direct cooling techniques. 
Here, 2DES would reveal the individual reaction and relaxation pathways by directly connecting the initial and final states and possible intermediate states. 

During the review process, we became aware of a parallel work by Hebin Li and co-workers on 2DES of a Rb MOT~\cite{Liang.20221018}.

\begin{backmatter}

\bmsection{Funding}
H2020 European Research Council (694965); Deutsche Forschungsgemeinschaft (RTG 2717); Fonds der Chemischen Industrie.

\bmsection{Acknowledgments}
K.D. was supported by the Fonds der Chemischen Industrie (Liebig Fellowship).
T.S. was supported by the Landesgraduiertenförderung Baden-Württemberg (completion scholarship).
We thank Tom Gallagher (University of Virginia) for stimulating discussions.

\bmsection{Disclosures} The authors declare no conflicts of interest.

\bmsection{Data availability}
Data underlying the results presented in this paper are not publicly available at this time but may be obtained from the authors upon reasonable request.

\end{backmatter}

\bibliography{references}


\end{document}